\documentclass[
aps,
prc,
showpacs,
twocolumn,
superscriptaddress,
nofootinbib,
floatfix]
{revtex4-1}
\usepackage{graphicx,
longtable,
slashed,
hyperref,
bm,url,
braket,
xcolor,
xspace
}

\usepackage[normalem]{ulem}

\newcommand{\bbz}{$0\nu\beta\beta$\xspace}
\newcommand{\bbt}{$2\nu\beta\beta$\xspace}

\begin{document}

\title{Chiral Two-Body Currents and Neutrinoless Double-Beta Decay in the QRPA}

\author{J. Engel}
\affiliation{Department of Physics and Astronomy, University of North Carolina,
Chapel Hill, NC, 27516-3255, USA} \email{engelj@physics.unc.edu}
\author{F. \v{S}imkovic}
\affiliation{Boboliubov Laboratory of Theoretical Physics, JINR 141980 Dubna,
Russia}
\affiliation{Department of Nuclear Physics and Biophysics, Comenius
University, Mlynsk\'{a} dolina F1, SK-842 48
Bratislava, Slovakia}
\affiliation{Institute of Experimental and Applied Physics, Czech Technical
University in Prague, 128-00 Prague, Czech Republic}
\author{P. Vogel}
\affiliation{Kellogg Radiation Laboratory, California Institute of Technology,
Pasadena, CA 91125, USA} 

\begin{abstract}
We test the effects of an approximate treatment of two-body contributions to
the axial-vector current on the QRPA matrix elements for neutrinoless
double-beta decay in a range of isotopes.  The form and strength of the
two-body terms come from chiral effective-field theory.  The two-body currents
typically reduce the matrix elements by about 20\%, not as much as in
shell-model calculations.  One reason for the difference is that standard
practice in the QRPA is to adjust the strength of the isoscalar pairing
interaction to reproduce two-neutrino double-beta decay lifetimes.  Another may
be the larger QRPA single-particle space.  Whatever the reasons, the effects on
neutrinoless decay are significantly less than those on two-neutrino decay,
both in the shell model and the QRPA.
\end{abstract}
 \date{\today}
\maketitle

\section{Introduction}
\label{sec:introduction}

The observation of neutrinoless double-beta (\bbz) decay would mean that
neutrinos are Majorana particles.  It would also tell us the overall neutrino
mass scale if the nuclear matrix elements that help govern the decay could be
calculated with sufficient accuracy.  At present, the matrix elements from
reasonable calculations differ from one another by up to factors of three.  The
true uncertainty might be larger if there is physics that none of the
calculations capture. 

One familiar source of uncertainty is the way in which the axial-vector
coupling constant, in the parlance of nuclear physicists, is ``renormalized in
medium.'' The renormalization has several apparent sources, not all of which
are directly connected to the weak interactions.  Truncation of the
many-nucleon Hilbert space, for instance, appears to reduce the matrix elements
of spin operators, whether or not they stem from weak interactions.  But a
separate source affects weak currents themselves: many-body operators that
arise because nucleons are effective degrees of freedom.  The many-body
currents reduce matrix elements as well, though by amounts that are still in
dispute. 

Here, we focus on two-body currents in a very particular framework: chiral
effective field theory ($\chi$EFT) in combination with the Quasiparticle Random
Phase Approximation (QRPA).  We build on several published papers.  Refs.\
\cite{n3lo} and \cite{epl05} extract coefficients of the $\chi$EFT interaction,
and Refs.\ \cite{par03} and \cite{gaz09} present the form of the $\chi$EFT one-
and two-body weak currents that follow.  Refs.\ \cite{men11} and \cite{klo13}
use an isospin-symmetric Fermi gas model to substitute approximate effective
one-body current operators for the complicated two-body operators, and use the
renormalized one-body current to correct shell-model calculations of
double-beta decay (in a particular limit that we describe later) \cite{men11}
and WIMP-nucleus scattering (in more generality) \cite{klo13}.  We will use the
effective one-body operators from both references to calculate double-beta
decay in the QRPA. 

There is some reason, before beginning, to believe that the effects of two-body
currents will be smaller in the QRPA, as usually applied, than in the shell
model.  The correlations in the QRPA are simpler than in the shell-model and in
recent years QRPA practitioners have compensated by fitting a parameter in the
interaction --- the isoscalar particle-particle interaction --- to reproduce
measured two-neutrino double-beta (\bbt) decay rates.  Addition to the axial
current operator will be compensated to retain the correct \bbt matrix element,
and the compensation should carry over, in some measure at least, to \bbz
decay.  We shall see below the degree to which that occurs.

The remainder of this paper is structured as follows: Section \ref{sec:methods}
describes the ingredients of our calculation, including the new two-body
currents.  Section \ref{sec:results} displays our results for \bbz matrix
elements in a wide range of isotopes.  Section \ref{sec:conclusion} is a
conclusion.

\section{Methods}
\label{sec:methods}

\subsection{One- and Two-body Currents}
 
In $\chi$EFT, interactions and currents for nucleon and pion degrees of freedom
are expanded in powers of momentum transfer $p$ divided by a breakdown scale
$\Lambda_\chi \approx 500$ MeV.  Following Ref.\ \cite{men11}, we equate
$\mathcal{O}(p/m)$, where $m$ is the nucleon mass, with
$\mathcal{O}((p/\Lambda_\chi)^2)$.  The interactions and currents should be
derived consistently, either through fitting or by matching onto the
predictions of QCD.  We will not use a $\chi$EFT interaction, but can still do
an approximately correct calculation by using $\chi$EFT currents.  Of course
those currents will comprise operators for three, four, \ldots nucleons and
there is no guarantee that the higher-order terms in the chiral expansion that
generates these operators will be small in a many-body system. Truncating the
expansion for the interaction at low order yields reasonable results, however,
and it is worth exploring a similar approximation in the currents.

In a non-relativistic framework, one can write a general one-body current in
the form
\begin{eqnarray} 
\label{eq:cur1}
J^{0\dag}(\mathbf{r}) &=& \sum_{i=1}^A  J^0_{i,1b} \delta
(\mathbf{r}-\mathbf{r}_i) \tau_i^+ \\
\label{eq:cur2}
\mathbf{J}^\dag(\mathbf{r}) &=& \sum_{i=1}^A \mathbf{J}_{i,1b} 
\delta(\mathbf{r}-\mathbf{r}_i)  \tau_i^+ \,. \nonumber
\end{eqnarray}
where an operator with subscript $i$ acts only on the $i^{\rm th}$ nucleon and
$\tau^+$ changes a neutron into a proton.  To third order in the counting, the
one-body charge-changing weak current operators can be written as \cite{men11}
\begin{eqnarray}
\label{eq:one-body-1}
J^0_{i,1b}  &=&
 g_V(p^2) - g_A(0)
\frac{\mathbf{P}\cdot \boldsymbol{\sigma}_i}{2m}  
+ g_P(p^2)\frac{E\boldsymbol{\sigma}_i\cdot \mathbf{p}}{2m} \,,  \\
\label{eq:one-body-2}
{\bf J}_{i,1b}  &=& \left(- g_A(p^2) \boldsymbol{\sigma}_i +g_P(p^2)
\frac{\mathbf{p}(\boldsymbol{\sigma}_i\cdot \mathbf{p})}{2m}\right.\\
&-& \left.i (g_M(0) + g_V(0)) \frac{\boldsymbol{\sigma} \times
\mathbf{p}}{2m} + g_V(0) \frac{\mathbf{P}}{2m} \right)  \,, \nonumber
\end{eqnarray}
where $\mathbf{p}=\mathbf{p}_i-\mathbf{p'}_i$,
$\mathbf{P}=\mathbf{p}_i+\mathbf{p'}_i$, $E=E_i-E_i'$, and, to the same order
in the chiral expansion,
\begin{eqnarray}
\label{eq:ch-cur2}
g_V(p^2) &=& 1-2\frac{p^2}{(850 \textrm{MeV})^2}\,, \\ 
g_A(p^2) &=& g_A(0)\left(1-2\frac{p^2}{(1040 \textrm{MeV})^2}\right) \,, \quad
g_A(0)=1.27 \,, \nonumber \\
g_P(p^2) &=& 2 \frac{g_{\pi p n} F_\pi}{p^2+m_\pi^2}-\frac{4mg_A(0)}{1040
\textrm{MeV}^2}\,, \qquad g_M(0) = 3.70 \nonumber
\end{eqnarray}
with $F_\pi=92.4$ MeV, $m_\pi=138.04$ MeV, and $g_{\pi p n} = 13.05$.  In all
these expressions $\mathbf{p}_i$ and $\mathbf{p}'_i$ stand for
$-i\mathbf{\nabla}_i$ acting on the left and right of the delta functions in
Eqs.\ (\ref{eq:cur1}) and (\ref{eq:cur2}).

As mentioned in the introduction, we will use two separate approximations
schemes for two-body currents, one presented in Ref.\ \cite{men11} for \bbz
matrix elements in the shell model and an improved version presented by the
same group in a paper on spin-dependent WIMP-nucleus scattering \cite{klo13}.
The first approximation scheme neglects the difference between the momentum
transfers to the two nucleons.  We use it nonetheless because it is the only
scheme applied so far to double beta decay and we wish to compare our results
with those of Ref.\ \cite{men11}.  The two schemes will turn out to yield only
minor differences.

Both schemes involve an effective correction to the one-body current through
the assumption that one of the two nucleons in the two-body current lies in a
spin-and-isospin symmetric core.  The resulting approximation is crude but
probably reasonable.  Ref.\ \cite{men11} neglects tensor-like terms in the
current, leading to a renormalization of $g_A$ but not $g_P$.  Ref.\
\cite{klo13} does a more complete calculation that leads to a separate
renormalization of $g_P$.  Here we write explicitly only the effective current
of \cite{men11}:
\begin{eqnarray}
\label{eq:two-body-eff}
\bra{\mathbf{p_i}}\mathbf{J}^{\rm eff}_{i,2b}(\mathbf{r})
\ket{\mathbf{p_i'}}&& =- g_A(p^2) \boldsymbol{\sigma}_i \left( \frac{\rho}{F_\pi^2} \left[ \frac{c_D}{g_a
\Lambda_\chi} + \frac{2}{3}c_3\frac{p^2}{4m\pi^2+p^2} \right. \right. 
\nonumber \\
&& \hspace*{-.3cm} \left. \left. + I(\rho,P) \left( \frac{1}{3}(2c_4-c_3) + \frac{1}{6m}
 \right) \right] \right) e^{-i\mathbf{p}\mathbf{r}} \,. 
\end{eqnarray}
with
\begin{eqnarray}
\label{eq:integral}
I(\rho,&P)& = 1 -\frac{3m_\pi^2}{2k_F^2}+\frac{3m_\pi^3}{2k_F^3}\textrm{acot}
\left[ \frac{m_\pi^2+\frac{P^2}{4}-k_F^2}{2m_\pi k_F} \right] \\
&+& \frac{3m_\pi^2}{4k_F^3P}\left( k_F^2+m_\pi^2-\frac{P^2}{4} \right)
\textrm{ln}\left[ \frac{m_\pi^2+(k_F-\frac{P}{2})^2}{m_\pi^2+(k_F+
\frac{P}{2})^2} \right] \,. \nonumber
\end{eqnarray}
In these equations $k_F$ is the Fermi momentum and $P$ is the center-of-mass
momentum of the decaying nucleons, which can be set to zero without altering
$I(\rho,P)$ significantly \cite{men11}.  The constants $c_3$, $c_4$, and $c_D$
are the $\chi$EFT parameters, fit to data in light nuclei.  Their values
depend on how the fit is carried out.

The above can be captured by defining new \emph{effective} one-body current
operators $\mathcal{J}^{\mu\dag}$ as the operators $J^{\mu\dag}$ from Eq.\
(\ref{eq:cur1}) but with the factor $g_A(p^2)$ multiplying
$\boldsymbol{\sigma}_i$ in Eq.\ (\ref{eq:one-body-2}) replaced by an effective
coupling $g_A^{\rm eff}(p^2)$, given by
\begin{eqnarray}
\label{eq:gaef}
g_A^{\rm eff}(p^2) &=& g_A(p^2) \left( 1-\frac{\rho}{F_\pi^2} \left[ \frac{c_D}{g_a
\Lambda_\chi} + \frac{2}{3}c_3\frac{p^2}{4m_\pi^2+p^2} \right. \right. \nonumber \\
 &&\left. \left. + I(\rho,0) \left( \frac{1}{3}(2c_4-c_3) + \frac{1}{6m}
 \right) \right] \right) \,, 
\end{eqnarray}
where $P$ has been set to zero.

The treatment of WIMP scattering in Ref.\ \cite{klo13} is more complete and
involves much longer expressions.  We refer the reader there for details on the
renormalization of both $g_P$ and $g_A$.

\subsection{Decay Matrix Elements}

The two kinds of double-beta decay --- \bbt and \bbz --- transfer very
different amounts of (virtual) momentum among nucleons.  Two-neutrino decay is
simply two successive virtual beta decays, with very little momentum transfer.
Its matrix element can be written with excellent accuracy as
\begin{equation}
\label{eq:2nume}
M'^{2\nu} = \left(\frac{g_A^{\rm eff}(0)}{g_A(0)}\right)^2 \sum_{N,i,j} \frac{\bra{F} \boldsymbol{\sigma}_i \tau_i^+ \ket{N} \cdot \bra{N}
\boldsymbol{\sigma}_j \tau_j^+ \ket{I}}{E_N-\frac{E_I-E_F}{2}} \,,
\end{equation}
where the $\ket{N}$ are states in the intermediate nucleus with energy $E_N$,
and $\ket{I}$ and $\ket{F}$ are the initial and final nuclear ground states,
with energies $E_I$ and $E_F$.  This matrix element and the neutrinoless
version to follow differ from the unprimed $M^{2\nu}$ (and $M^{0\nu}$) used
elsewhere in that $g_A$ is always set to $g_A(0) = 1.27$ (and not to some
effective value) in the phase space factor multiplying the matrix element, so
that all effects of $g_A$ modification are in the matrix element itself.

Neutrinoless decay, in contrast to its two-neutrino counterpart, creates a
virtual neutrino that typically carries about 100 MeV of momentum.  The
expression for its matrix element involves an integral over all neutrino
momenta:
\begin{eqnarray}
\label{eq:0nume}
M^{'0\nu} &=& \frac{R}{2 \pi^2 g_A(0)^2} \sum_{N} \!\! 
 \int \! d^3x \, d^3y \,  d^3p \\
&\times& e^{i\mathbf{p}\cdot (\mathbf{x}-\mathbf{y})} \frac{\bra{F} 
\mathcal{J}^{\mu\dag}(\mathbf{x})
\ket{N}\bra{N}
\mathcal{J}^\dag_\nu (\mathbf{y}) \ket{I}}{p (p+E_N-\frac{E_I-E_F}{2})}  \,, 
\nonumber
\end{eqnarray}
where $R$ is the nuclear radius, inserted to make the matrix element
dimensionless.  Details on the evaluation of this still rather abstract
expression appear, e.g., in Refs.\ \cite{sim99} and \cite{sim08}.  The
important point is that the \bbz matrix element depends on $g_A^{\rm eff}(p^2)$
(and in the formulation of Ref.\ \cite{klo13} on $g_P^{\rm eff}(p^2)$ as well)
because of the two current operators $\mathcal{J}^{\mu\dag}$, defined just above
Eq.\ (\ref{eq:gaef}), and the integral over momentum.  

\begin{table*}[h] 
\caption{ The $0\nu\beta\beta$ matrix element ${M'}^{0\nu}_{}$ with one- and
two-body nucleon current operators from the text and the
Argonne-V18-G-matrix-based QRPA.  We use several sets of values for the
$\chi$EFT parameters and two nuclear densities, and both the simplest and more
complete versions of the effective one-body current, from Refs.\ \cite{men11}
and \cite{klo13} respectively.  $\langle{M'}^{0\nu}_{}\rangle$ is the matrix
element averaged over these possibilities; its variance is in parentheses.  The
columns labeled $a$ through $d$ correspond to different EFT-parameter choices
(defined in Ref.\ \cite{men11}) and nuclear-density choices.  These choices are a: EGM+$\delta c_i$,
$\rho$= 0.10 fm$^{-3}$; b: EGM+$\delta c_i$, $\rho$= 0.12 fm$^{-3}$; c: EM,
$\rho$= 0.10 fm$^{-3}$; d: EM, $\rho$= 0.12 fm$^{-3}$.  The last column
contains the percent suppression $\varepsilon$ of
$\langle{M'}^{0\nu}_{}\rangle$ with respect to the value without two-body
currents (displayed in the first column).}
\label{tab:all1}    
\centering 
\begin{tabular}{lcccccccccccccc}
\hline \hline 
nucleus & ${M'}^{0\nu}_{}$ & & \multicolumn{9}{c}{${M'}^{0\nu}_{}$ (2bc)} & &
$\langle{M'}^{0\nu}_{}\rangle$ & $\varepsilon$ \\ \cline{4-12} 
	 & 1bc & & \multicolumn{4}{c}{param. of Ref. \cite{men11}} & &
	 \multicolumn{4}{c}{param. of Ref. \cite{klo13}}  & & with & \\
\cline{4-7} \cline{9-12} & & & a & b & c & d & & a & b & c & d & & quenching &
[\%] \\ 
\hline\hline
${^{48}Ca}$   & 0.684 & & 0.641 & 0.629 & 0.580 & 0.558  & & 0.637 & 0.637 & 0.596 & 0.592 & & 0.61(0.03)  & 11 \\ 
 ${^{76}Ge}$  & 5.915  & & 5.121  & 4.932 & 4.369 & 4.084  & & 5.050 & 4.914 & 4.412  & 4.206  & & 4.64(0.41) & 22 \\ 
 ${^{82}Se}$  & 5.313  & & 4.570  & 4.393 & 3.863 & 3.583  & & 4.506 & 4.378 & 3.906  & 3.701  & & 4.11(0.39) & 23 \\
 ${^{96}Zr}$  & 3.224  & & 2.999  & 2.913 & 2.636 & 2.506  & & 2.946 & 2.894 & 2.651  & 2.573  & & 2.76(0.19) & 14 \\ 
 ${^{100}Mo}$  & 6.287  & & 5.552  & 5.370 & 4.801 & 4.510  & & 5.437 & 5.314 & 4.813  & 4.618  & & 5.05(0.41) & 20  \\ 
 ${^{110}Pd}$  & 6.575  & & 5.795  & 5.607  & 5.037 & 4.758  & & 5.673 & 5.540 & 5.030 & 4.833  & & 5.28(0.41) & 20 \\ 
 ${^{116}Cd}$  & 4.485  & & 3.894 & 3.754 & 3.342 & 3.127  & & 3.812 & 3.701 & 3.331 & 3.126 & & 3.51(0.31) & 22 \\ 
 ${^{124}Sn}$  & 3.974  & & 3.599 & 3.511 & 3.231 & 3.118  & & 3.521 & 3.464 & 3.211 & 3.143 & & 3.35(0.19) & 16 \\ 
 ${^{130}Te}$  & 4.610  & & 4.031  & 3.890 & 3.445 & 3.216  & & 3.949 & 3.855 & 3.465  & 3.313  & & 3.65(0.32) & 21 \\ 
 ${^{136}Xe}$  & 2.570  & & 2.249  & 2.169 & 1.920 & 1.791  & & 2.190 & 2.136 & 1.915  & 1.829  & & 2.02(0.18) & 21 \\ 
\hline\hline
\end{tabular}  
\end{table*}    
\begin{table*}[!h] 
\caption{The same as Table \ref{tab:all1}, but for the CD-Bonn interaction
instead of the Argonne V18 interaction }
\label{tab:all2}    
\renewcommand{\arraystretch}{1.1} \centering 
\begin{tabular}{lcccccccccccccc}
\hline \hline 
nucleus & ${M'}^{0\nu}_{}$ & & \multicolumn{9}{c}{${M'}^{0\nu}_{}$ (2bc)} & &
$\langle{M'}^{0\nu}_{}\rangle$ & $\varepsilon$ \\ \cline{4-12} 
	 & 1bc & & \multicolumn{4}{c}{param. of Ref. \cite{men11}} & &
	 \multicolumn{4}{c}{param. of Ref. \cite{klo13}}  & & with & \\
\cline{4-7} \cline{9-12} & & & a & b & c & d & & a & b & c & d & & quenching &
[\%] \\ 
\hline\hline
  ${^{48}Ca}$               &  0.649  & & 0.615 & 0.605 & 0.561 & 0.542  & & 0.606 & 0.606 & 0.570  & 0.569 & & 0.58(0.03)  & 10 \\ 
  ${^{76}Ge}$              &   5.849  & & 5.086  & 4.904 & 4.356 & 4.082  & & 4.990 & 4.858 & 4.371  & 4.175  & & 4.60(0.40) & 21\\ 
  ${^{82}Se}$              & 5.255  & & 4.538  & 4.366 & 3.848 & 3.577  & & 4.453 & 4.327 & 3.867  & 3.669  & & 4.08(0.38) & 22 \\
  ${^{96}Zr}$              & 3.144  & & 2.953  & 2.872 & 2.608 & 2.485  & & 2.883 & 2.835 & 2.603  & 2.532  & & 2.72(0.18) & 12 \\ 
  ${^{100}Mo}$              & 6.164  & & 5.469  & 5.295  & 4.747 & 4.469  & & 5.326 & 5.208 & 4.726  & 4.542  & & 4.97(0.39) & 19 \\ 
 ${^{110}Pd}$               & 6.532  & & 5.772  & 5.589  & 5.029 & 4.758  & & 5.629 & 5.497 &  4.998 & 4.806  & & 5.26(0.40) & 19 \\ 
  ${^{116}Cd}$              & 4.474  & & 3.888  & 3.749 & 3.338 & 3.125  & & 3.796 & 3.685 & 3.317 & 3.149 & & 3.51(0.31) & 22  \\ 
  ${^{124}Sn}$              & 4.024  & & 3.646 & 3.556 & 3.273 & 3.158  & & 3.553 & 3.494 & 3.239 & 3.170 & & 3.29(0.20) & 16 \\ 
  ${^{130}Te}$              & 4.642  & &  4.063 & 3.921  & 3.473 & 3.242  & & 3.958 & 3.861 & 3.468  & 3.313  & & 3.66(0.32) & 21 \\ 
  ${^{136}Xe}$              & 2.602  & &  2.276 & 2.196  & 1.943 & 1.812  & & 2.206 & 2.149 & 1.926  & 1.837  & & 2.04(0.18) & 21 \\ 
\hline\hline
\end{tabular}  
\end{table*}

\section{Results}
\label{sec:results}
The values of the parameters $c_3$, $c_4$, and $c_D$ come from fits to data in
systems with very few nucleons.  They depend on details of the fitting
procedure; for this reason Ref.\ \cite{men11} gives several sets of possible
values.  It also evaluates the effective one-body current for a range of
Fermi-gas densities $\rho$ (the gas represents the nuclear core) because the
nuclear density, though roughly constant in the nuclear interior, is not
exactly so.  As a result, it finds a range of final shell-model \bbz matrix
elements, with the correct one probably somewhere within the range.  Here we
will use $c$ parameters and densities at the extremes of the reasonable range
to set probable upper and lower limits on the effects of two-body currents.  

\begin{figure}[!t]
\centering
\includegraphics[width=1.1\columnwidth]{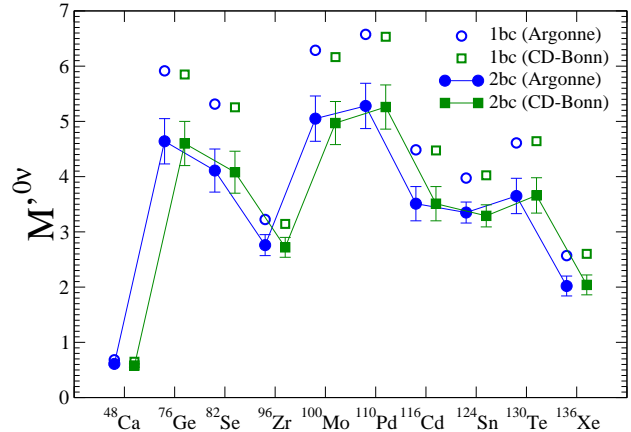}
\caption{(Color online) \label{fig:all} Nuclear matrix elements $M'^{0\nu}$ for
all the nuclei considered here.  The empty circles and squares represent the
results with the one-body current only, and the solid circles and squares the
average of the results with two-body currents included.  The error bars
represent the dispersion in those values (see text).  }
\end{figure}

Tables \ref{tab:all1} and \ref{tab:all2} present the results of our
calculations.  The headings $a,b,c$ and $d$ in these tables refer to various
prescriptions for fixing the $\chi$EFT parameters (see caption).  The last
column averages the quenching of the \bbz matrix element over these entries,
leading to a mean effect of about 20\%, either with the parameterization of
Ref.\ \cite{men11} or the more complete one in Ref.\ \cite{klo13}.  Fig.\
\ref{fig:all} summarizes the same results, comparing $M'^{0\nu}$ with one-body
and one-plus-two-body currents for all the nuclei we consider.  The degree of
quenching is noteworthy for two reasons.  First, the \bbz quenching is much
less than its \bbt counterpart, which with the same currents is closer to 40 or
50\%.  Second, it is noticeably less then the quenching of \bbz decay in the
shell model.

The minimum quenching from the set of choices in Ref.\ \cite{men11} occurs when
$c_3 = -3.2$, $c_4 = 8.6$, and $c_D = 0$ \cite{n3lo}, a combination we call EM
below (following Ref.\ \cite{men11}) while the maximum quenching corresponds to
$c_3 = -2.4, c_4 = 4.8$ and $c_D = 0.0$ \cite{epl05}, a combination we call
EGM+$\delta c_i$. (The units of the $c$ parameters are GeV$^{-1}$.)  These
choices result in values for $g_A^{\rm eff}(0)/g_A$ in a range 0.66 -- 0.85
that brackets the empirical value of that ratio derived from the analysis of
ordinary Gamow-Teller beta decay; see e.g.\ Refs.\  \cite{bro88} and
\cite{mar96}.  Ref.\  \cite{lisi} points out that single $\beta$ and
two-neutrino double $\beta$ decay observables can be described simultaneously
in the QRPA with $g_A^{\rm eff}(0)/g_A$ in that range, implying that two-body
currents can completely account for the renormalization of $g_A$.  On the other
hand, older meson-exchange models \cite{tow87} suggest that the effects of
many-body currents on allowed beta decay are small.  The source of the
disagreement between the strength of two body currents in $\chi$EFT and
exchange models is not completely clear to us.

There are several reasons for the first fact.  As Ref.\ \cite{men11} shows, the
degree of quenching decreases with increasing momentum transfer. An as we noted
earlier, \bbt decay involves almost no momentum transfer by the currents, while
\bbz decay involves momentum transfers that are typically about 100 MeV and
still contribute non-negligibly at several hundred MeV.  In addition, the \bbz
matrix element contains a Fermi part, for which we have assumed no quenching.
While this assumption may not be completely accurate, it is implied at low
momentum transfer by CVC.  The overall quenching of the vector current is
certain to be less than that of the axial-vector current. (In the results
listed in Tables \ref{tab:all1} and \ref{tab:all2} the Fermi matrix elements
are smaller than in some other calculations because the isovector
particle-particle interaction was adjusted as explained in Ref.\
\cite{simkovic13} to reflect isospin symmetry.)

Why is the QRPA \bbz quenching less than that in the shell model?  Part of the
reason, as we noted in the introduction, is that in the QRPA the strength of
the isoscalar pairing interaction, which we call $g_{pp}^{T=0}$, is adjusted to
reproduce the measured \bbt rate.  The suppression of \bbt decay by two-body
currents implies that the value of $g_{pp}^{T=0}$ is smaller than it would be
without those currents.  The smaller $g_{pp}^{T=0}$ in turn implies less
quenching for the \bbz matrix element.

\begin{figure}[b]
\centering
\includegraphics[width=.5\textwidth]{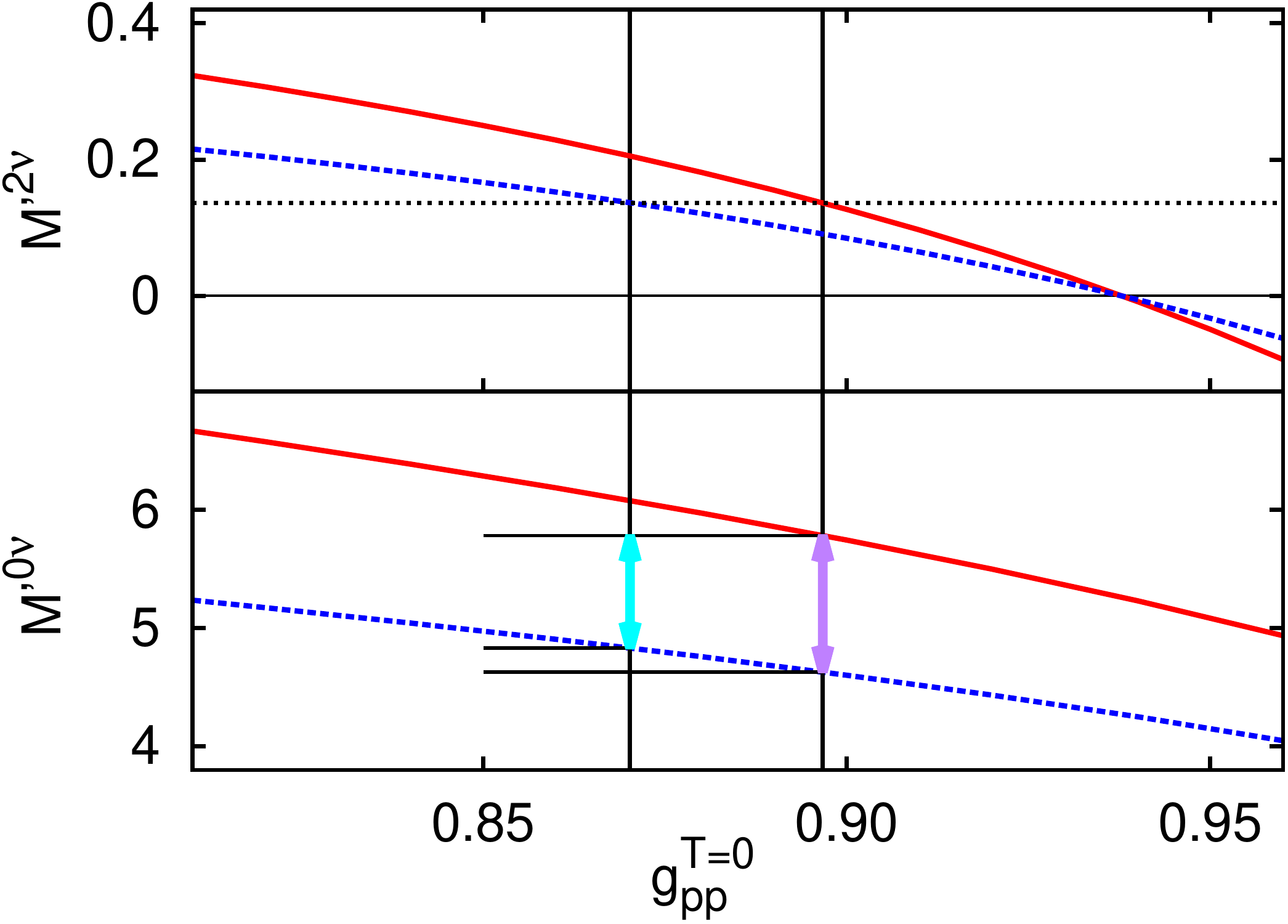}
\caption{(Color online) \label{fig:gpp} The quenching of \bbt and \bbz decay 
by two-body currents in $\chi$EFT.  Top:  $M'^{2\nu}$ vs.\ the $g_{pp}^{T=0}$,
the strength of isoscalar pairing. The solid (red) line is the unquenched
matrix element and the dashed (blue) line the matrix element with quenching
caused by two-body currents, with the parameterization EGM+$\delta c_i$ from
Ref.\ \cite{men11}.  The dotted black line is the measured matrix element
\cite{ago13} under the assumption that $g_A$ is unquenched.  The vertical lines
are the values of $g_{pp}^{T=0}$ that reproduce the measurement with and
without two-body currents.  Bottom:  The same, for $M'^{0\nu}$ (without a
measured value).  The long (purple) arrow represents the quenching when
$g_{pp}^{T=0}$ is not readjusted to reproduce \bbt decay.  The short (cyan)
arrow is the quenching when $g_{pp}^{T=0}$ is readjusted.}
\end{figure}

Figure \ref{fig:gpp} illustrates this idea.  The upper panel shows the \bbt
matrix element, with (solid red) and without (dashed blue) two-body currents.
The two vertical lines indicate the values of $g_{pp}^{T=0}$ needed to
reproduce the ``measured'' matrix element \cite{ago13}, defined as that which
gives the lifetime under the assumption that $g_A$ is unquenched. The value of
$g_{pp}^{T=0}$ that works with the two-body currents is smaller.  The lower
panel shows the consequences for \bbz decay.  The longer (purple) arrow
represents the quenching that would obtain if $g_{pp}^{T=0}$ were not adjusted
for the presence of the two-body currents (as is the case in the shell model,
where the interaction is fixed ahead of time).  The shorter arrow represents
the same quenching after adjusting $g_{pp}^{T=0}$.  The requirement that we
reproduce \bbt decay thus means that the \bbz matrix element is quenched
noticeably less than it would otherwise be.  

Another difference between the QRPA and the shell model is that the QRPA works
in a much larger single-particle space (at the price of working with only a
particular kind of correlation).  This larger space presumably means larger
contributions at high momentum transfer.  Since the quenching decreases with
momentum transfer, the contributions of the high-angular-momentum multipoles
are less affected by the two-body currents than their low-angular-momentum
counterparts.  The large QRPA model space therefore suggests that the quenching
of \bbz decay is less than it would be in a shell model calculation.  The size
of this effect, however, is hard to quantify.

\section{Discussion}
\label{sec:conclusion}

It is clear, in today's terminology, that some of the quenching of spin
operators in nuclei is due to the use of restricted model spaces and some to
many-body currents.  Model-space truncation can exclude strength that may be
pushed to high energies, and the omission of two-body currents leaves
delta-hole excitations, among other things, unaccounted for.  The question of
which effect is more important is still open.  If two-body currents are behind
most of the quenching, as recent fits of the $c$ parameters seem to suggest,
then \bbt decay is very likely more quenched than \bbz decay and existing
calculations of \bbz decay that don't include quenching are at least in the
right ballpark.  We've seen, under this assumption, that such is the case in
the QRPA, even a bit more than in the shell model.

It is still possible, as older meson-exchange models suggest \cite{tow87}, that
the effects of many-body currents are small.  In that event the quenching of
neutrinoless decay would be unrelated to the two-body currents and could be
similar in magnitude to the quenching of two-neutrino decay, a state of affairs
that would make \bbz experiments less sensitive to a Majorana neutrino mass
than we currently believe.  A strong argument that this state of affairs is
real, however, has yet to be presented.  It seems likely to us that the
quenching of \bbz matrix elements is around the size indicated by the
$\chi$EFT-plus-QRPA analysis carried out here.

\vspace*{-.2cm}
\begin{acknowledgments}

This work was supported by the U.S.\ Department of Energy through Contract No.\
DE-FG02-97ER41019. F. \v S.\ acknowledges the support by the VEGA Grant agency
of the Slovak Republic under the contract No.\ 1/0876/12 and by the Ministry of
Education, Youth and Sports of the Czech Republic under contract LM2011027.

\end{acknowledgments}

\end{document}